\documentclass{ws-procs975x65}

\begin{document}

\title{DETECTION OF EXOPLANETS IN M31 WITH PIXEL-LENSING: 
THE EVENT PA-99-N2 CASE}

\author{G. INGROSSO$^*$ and F. DE PAOLIS}
\address{Dipartimento di Fisica, Universit\`a del Salento and
INFN Sezione di Lecce, Italy, \\ 
$^*$E-mail: ingrosso@le.infn.it}

\author{S. CALCHI NOVATI}
\address{Dipartimento di Fisica, Universit\`a di Salerno and INFN
Sezione di Napoli, Italy}

\author{PH. JETZER}
\address{Institute for Theoretical Physics, University of  Z\"{u}rich, 
Switzerland}

\author{A. A. NUCITA}
\address{XMM-Newton Science Operations Centre, ESAC, ESA, Madrid, Spain}

\author{A. F. ZAKHAROV}
\address{ITEP, Moscow, and 
BLTP, JINR, Dubna, Russia}

\begin{abstract}
We show that exoplanets in the M31 galaxy may be detected with   
the pixel-lensing method by using telescopes making high
cadence observations of an ongoing microlensing event.  
Although the mean mass for detectable exoplanets  
is about $2~M_{\rm {J}}$, even small mass exoplanets 
($M_{\rm P} < 20~M_{\oplus}$) can cause significant 
deviations, which are observable with large telescopes.
We reanalysed the POINT-AGAPE microlensing event PA-99-N2. 
First, we test the robustness
of the binary lens conclusion for this light curve.
Second, we show that for such long duration and bright
microlensing events, the efficiency for finding
planetary-like deviations is strongly enhanced.
\end{abstract}


\bodymatter


$~~~~~~~~~~~~~~~~$ \\

\begin{figure}
\begin{center}
\includegraphics[width=0.75\textwidth]{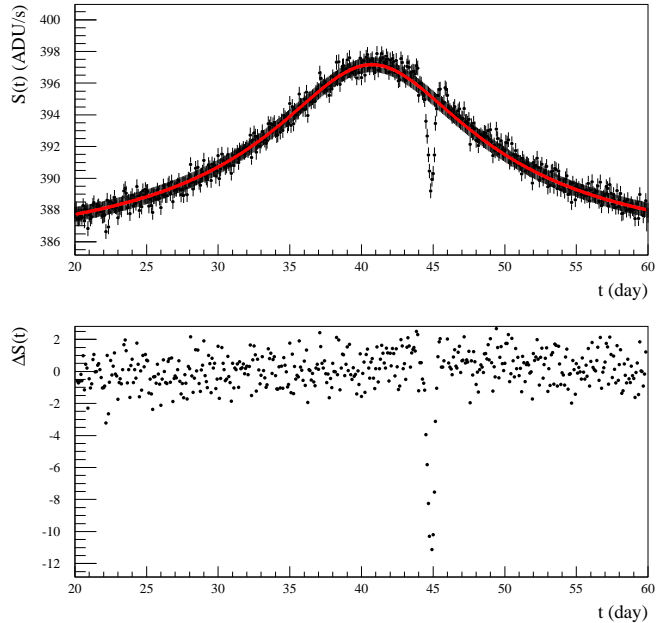}
\end{center}
\caption{Simulated II class event (points) and Paczy\`nki fit 
(thin solid line).
Some parameter values: 
$\rho/u_0=0.04$, 
$R_{\rm max}=24.0$ mag, 
$t_{1/2}=18.7$ day, 
$M_{ \rm P}=0.3~M_{\oplus}$,
$d_{\rm P}=3.8 $AU.}
\label{proceeding4}       
\end{figure}

The possibility to detect planets in pixel-lensing observations towards M31 
has been already explored earlier.
The analysis for planet detection, however, has been performed by using
a fixed configuration of the underlying Paczy\'{n}ski light curve.
Using a Monte Carlo (MC) approach we study the chances to 
detect exoplanets in M31, by considering the multi-dimensional 
space of parameters for both lensing and planetary systems \cite{mnras}. 
The method of residuals allows us to select the light curves that show 
detectable deviations with respect to the Paczy\'{n}ski shape (describing 
microlensing by single lens).
The advantage of the MC approach is that of allowing
a detailed characterisation of the sample of microlensing events 
for which the planetary deviations are more likely to be detected. 
We discriminate two classes of events (indicated by I and II), 
according to the ratio $\rho/u_0 >1$ or $\rho/u_0 <1$. 
The events of the I class have short durations 
($ \langle t_{1/2}\rangle \simeq 1.6$ day) and 
larger flux variations ($ \langle R_{\rm max}\rangle \simeq 20.6$ mag). 
In these events, planetary deviations are caused by the 
source trajectory crossing (in the lens plane) the central caustic region, 
close to the primary lens star.
The events of the II class have longer durations 
($\langle t_{1/2} \rangle =6.4$ day) and smaller flux variations  
($\langle R_{\rm max} \rangle=23.1$ mag). In this case 
planetary perturbations are (mainly) 
caused by the intersection of the source trajectory with the planetary 
caustic regions and may also appear at times  
far from the maximum amplification time $t_0$.
The fraction of I class events is about 5\% of all generated events. 
However, it turns out   
that the probability to have detectable planetary features in these events
(that however are rare) is higher. This happens since the crossing of the 
central caustic region is more probable in I class events (with $\rho/u_0>>1$).
On the contrary, the generated 
events of the II class are more numerous, but have 
a smaller probability to show detectable planetary features. 
An example of II class event is given in Fig. \ref{proceeding4}, 
where we see that also a small mass planet ($M_{ \rm P} = 0.3 ~ M_{\oplus}$) 
can cause detectable planetary deviations in events for which 
the finite size effects are small. 
In the simulated  event the geometry is such that the source 
trajectory is passing (in the lens plane) in the region
between the two planetary caustics,
where a deficit of amplification is present.
Other examples of light curves for I and II class events are given in paper
\cite{mnras}.

\begin{figure}
\begin{center}
\includegraphics[width=0.75\textwidth]{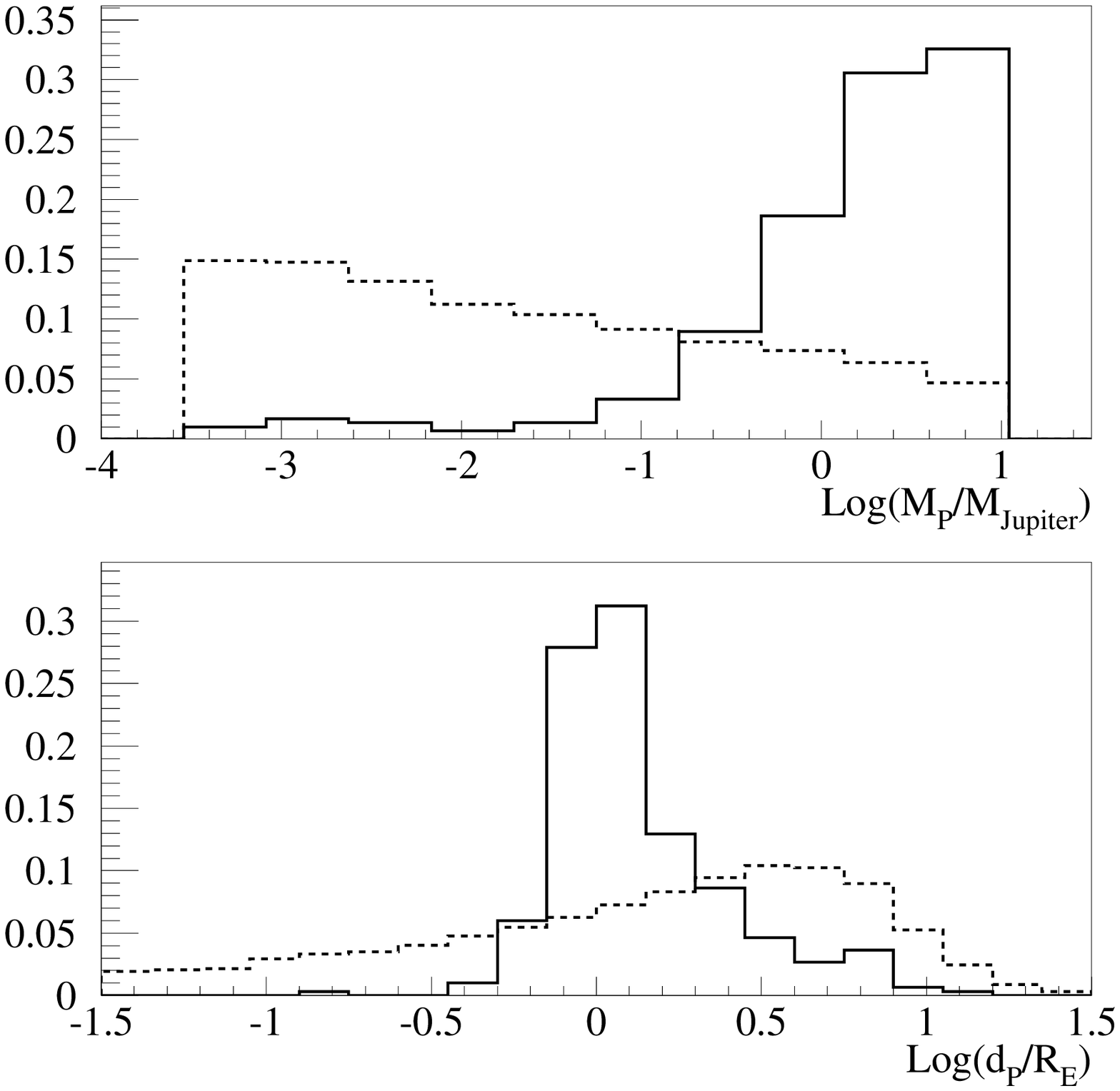}
\end{center}
\caption{Upper panel: distribution of the planet mass $M_{\rm P}$
for the events with detectable planetary deviations (solid line) 
and for the generated events (dashed line).
Bottom panel: distribution of $d_{\rm P}/R_{\rm E}$ for events as before.}
\label{parigi3}       
\end{figure}

Our analysis shows (see Fig. \ref{parigi3}, upper panel) that, 
in spite of the initial distributions, larger planetary masses 
lead to higher probability for the detection of planetary features 
in light curves. We also find that the planet detection is maximized 
when the planet-to-star separation $d_{\rm P}$ is inside the ``lensing zone''
(see Fig. \ref{parigi3}, bottom panel).
Typical duration of planetary perturbations is about 1.4 days. 
However, the number of significant planetary deviations on each light 
curve increases with increasing ratios $\rho/u_0$. So, the 
overall time scale for planetary deviations can  
increase up to 3 -- 4 days (for I class events).
This means that a reasonable time step for pixel-lensing 
observations aiming to detect planets in M31 is a few (4--6)  hours.
\begin{figure}
\includegraphics[width=0.95\textwidth]{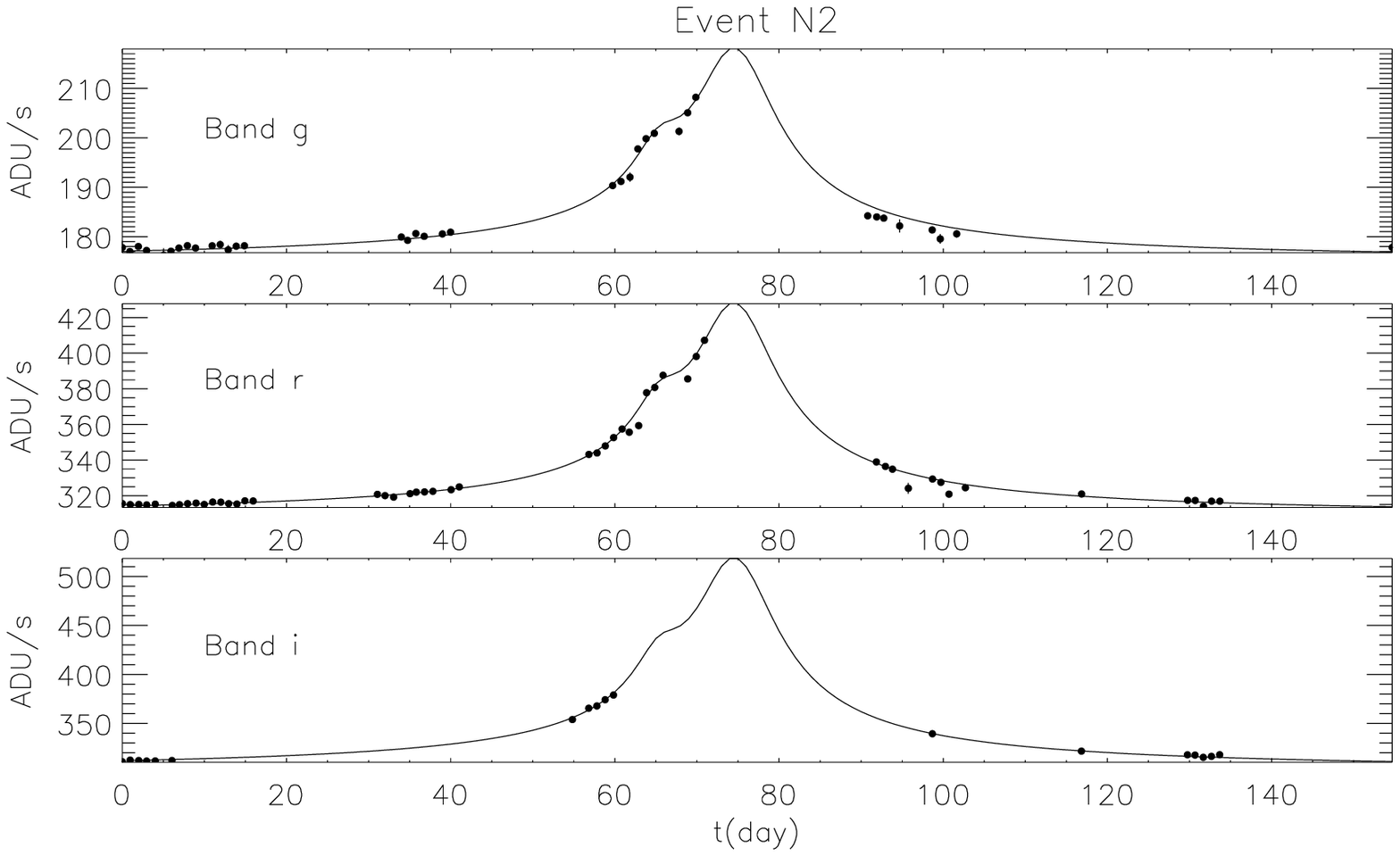}
\caption{The binary light curves corresponding to the C1 model \cite{an04} 
for g, r and  i bands.}
\label{n2-bin.eps}       
\end{figure}
Among the 25 pixel-lensing events 
discovered up to now \cite{review_novati}, 
the POINT-AGAPE collaboration  \cite{paulin03}
reported the detection of the event
PA-99-N2. This appears as a peculiar event because of 
its extreme brightness ($R_{\rm max} \simeq 19$ mag), long duration 
($t_{1/2} \simeq 24$ day)  and location, some 22 arcmin
away from the M31 center. Furthermore, as it has been
shown in a subsequent analysis \cite{an04}, the light curve
can be attributed to a secondary component orbiting the lens star,
because of the anomaly with respect to the Paczy\`nki shape.
In particular, An et al. \cite{an04} have evaluated the 
{\it a posteriori} probability distribution for the lens mass which 
results to be extremely broad: for source and lens disc objects, 
they report the lens mass range 0.02 -- 3.6 $M_{\odot}$ at 95\% confidence 
level. Together with the small binary 
lens mass ratio, $q \simeq 1.2 \times 10^{-2}$ for the best fit,
this puts the lens companion
well in the substellar range. Ingrosso et al. \cite{mnras}
have remarked that taking the likely value for the 
lens mass $\simeq 0.5~M_{\odot}$ for a disc lens, the lens companion would be 
a $\simeq 6.34~M_{\rm J}$ object. 
This would make of the PA-99-N2 lens companion
the first exoplanet discovered in M31. Furthermore, we had analysed 
the PA-99-N2 event within the framework of our simulation scheme,
showing in particular that its parameters 
nicely fall in the expected range for II class events \cite{mnras}. 
We further analyse this event \cite{grg09}. 
First, starting from the observational data
(courtesy of the POINT-AGAPE collaboration),
we test the robustness of the binary-lens best fit
solution.  Second, we address the question 
of the efficiency for finding binary-like
deviations for such bright and long duration events.
To verify the robustness of the binary-lens fit solution, 
the first test is to add a Gaussian noise to the best fit  
of the observed light curve to verify if a single lens model 
with noise can reproduce the observational data. 
Similarly, we have taken the best fit binary model (named C1 
in Table 1 of An et al. \cite{an04}) and realised more than one thousand 
models by adding Gaussian noise and letting the parameters to vary by 
at most 20\%. For each of the so obtained light curves (107 data points),
we have verified that the $\chi^2$ values for single lens models (8 parameters)
are greater (by a factor of about 3) with respect to those corresponding to 
binary fit models (11 parameters). We also find that the two $\chi^2$
distributions 
are clearly separated, 
which implies that the best single lens fit is much worst than any of the 
binary lens models. From this we conclude that the binary fit is robust and 
that the observed light curve cannot be fitted by any single lens model
with random noise.

A further point to be stressed is the following: 
we mentioned  that the probability 
to detect an  exoplanet in pixel-lensing observations
towards M31 is rather small, even with large telescopes \cite{mnras}.
Therefore, the question which arises now is the following: 
what is the chance of finding a planetary feature in an event as PA-99-N2? 
The basic answer can be found looking at the characteristics
of the underlying microlensing event:
this is, at the same time, an extremely
bright and long duration event. In fact, as we now show,
this strongly enhances the probability for finding
deviations to the single lens shape. 
To this purpose we perform a MC simulation
where we fix the single lens parameters $R_{\rm max}$ and $t_{1/2}$ 
to those of PA-99-N2 and we let vary the binary ones
(planet-to-star mass ratio, planetary distance and 
orientation of the source trajectory with respect to the binary axis) 
for an observational setup fixed to reproduce
these POINT-AGAPE observations.
For the selected  planetary mass range we find an increase
of the average efficiency up to 6\%, to be compared with 
0.6\% (see Table 3 in paper \cite{mnras}) 
for events without any constraints on 
$R_{\rm max}$ and $t_{1/2}$. 
The efficiency for finding binary-like deviations in the simulated light 
curves increases significantly with the lens-companion mass value.
In particular, for the restricted range $1-10~M_{\rm J}$
the average efficiency rises up to 27\% of the generated events.
We have restricted our attention to the planetary-mass range
for the lens companion, but of course
we expect the efficiency still rises
when we include the brown-dwarf mass range 
(as in fact it was allowed by the analysis \cite{an04}). 

This further analysis shows that the probability to find exoplanets
in M31, at least in the Jupiter mass range, with the pixel-lensing technique
is not so small as previously thought and thus 
that it is now mature for implementation. Clearly, the use of telescopes
with wide fields of view is needed in order to 
get a large enough number of events and thus to get sufficient statistics.
Actually, a detection of planetary features in  pixel-lensing 
events may be done also with telescopes of small field of view, if early 
warning systems (similarly to the OGLE one) will be in operation. 
Finally,  we remark that gravitational microlensing is a very efficient
method for discovering exoplanets around habitable zone, namely 
planetary systems with Earth-like masses and distances of about AU.

\bibliographystyle{ws-procs95x65}
\bibliography{ws-pro-sample}
\end{document}